\documentclass[pss,fleqn]{w-art}
\usepackage{times}
\usepackage{w-thm}
\theoremstyle{plain}

\theoremstyle{definition}

\usepackage[]{graphicx}
\chardef\bslash=`\\ 

\begin{document}
\DOIsuffix{theDOIsuffix}
\Volume{XX}
\Issue{1}
\Month{01}
\Year{2003}
\pagespan{1}{}
\Receiveddate{15 November 2003}
\Reviseddate{30 November 2003}
\Accepteddate{2 December 2003}
\Dateposted{3 December 2003}



\title[]{Localization in 2D quantum percolation}


\author[Igor Trav\v enec]{ Igor Trav\v enec \footnote{Corresponding author: e-mail: {\sf fyzitrav@savba.sk}}\inst{1}} \address[\inst{1}]{ Institute of Physics, Slovak Academy of Sciences, D\'ubravsk\'a cesta 9, SK - 84511 Bratislava, Slovakia}
\begin{abstract}
Quantum site percolation as a limiting case of binary alloy is studied numerically in 2D within the tight-binding model. We address the transport properties in all regimes - ballistic, diffusive (metallic), localized and crossover between the latter two. Special attention is given to the region close to the conduction band center, but even there the Anderson localization persists, without signs of metal - insulator transition. We found standard localization for sufficiently large samples. For smaller systems, novel partial quantization of Landauer conductances, i. e. most values close to small integers in arbitrary units is observed at band center. The crossover types of conductance distributions (outside the band center) are found to be similar to systems with corrugated surfaces. Universal conductance fluctuations in metallic regime are shown to approach the known, theoretically predicted value. The resonances in localized regime are Pendry necklaces. We tested Pendry's conjecture on the probability of such rare conducting samples and it proved consistent with our numerical results.

\end{abstract}
\maketitle                   

\section{Introduction}

The quantum percolation model exhibits Anderson localization and it was reviewed e. g. in Ref. \cite{mds}. We believe that it belongs to the same universality class as the Anderson model, thus it undergoes no metal - insulator transition (MIT) in 2D, i. e. all states become localized for any positive disorder \cite{sg,al} and large enough samples. We will regard square, or generally hypercubic disordered samples with $L^d$ sites, coupled to perfect semi-infinite leads. $d$ is dimensionality. Two characteristic lengths can be introduced, the mean free path $l$ and the localization length $\xi$. For finite samples and small disorder, with system size $L$ significantly smaller than both $l$ and $\xi$ we get the ballistic regime and roughly for $l < L < \xi$ the (quasi)metallic one, with mean Landauer conductance $\langle g\rangle \propto L^{d-1}$ and $\langle g\rangle \propto L^{d-2}$, respectively. In the localized regime $L > \xi$ and $\langle g\rangle \propto \exp(-L/\xi)$. At MIT point, characterized by critical amount of disorder, the localization length $\xi$ diverges and it remains infinite in full metal. Such MIT surely exists for $d>2$. In quasi-metallic regime $\xi$ is large, but finite, typically for $d \le 2$. The absence of MIT is known also in 2D Anderson model with orthogonal symmetry, i. e. neither with magnetic field nor spin-orbit interaction. The same was numerically checked again \cite{hks} for 2D quantum percolation. On contrary, in Ref. \cite{dca} and quite recently in a paper \cite{im}, MIT presence at 2D quantum percolation was reported. We aim to support the former statement that no indications of MIT are found when calculating the dependence of Landauer conductance on system size and other model parameters numerically.

Concerning experiments, in a recent paper \cite{clg} the authors report direct confirmation of MIT for 3D Anderson model, including correct critical exponent, characterizing the divergence of $\xi$. These measurements were performed on atomic matter waves, cooled Cs atoms kicked by laser pulses. Such experiment rules out problems of decoherence or many-body interactions, thus it is well described by our model.
 
However with real mesoscopic samples, experimenters often observe the MIT on 2D structures, see e. g. the review \cite{aks}. But even in our non-interacting, or one-electron models, switching the magnetic field or spin - orbit interaction already induces the MIT in 2D. At least the metallic regime seemed to be well understood and widely accepted. However, later measurements \cite{mw} on quasi-1D gold wires showed, than even in this regime the third cumulant of $g$ was non-zero, thus the distribution of conductances differs from the expected Gaussian. This can be explained e. g. by electron - electron interaction. One-particle models predict single parameter scaling and Gaussian distributions within the conduction band, though with possible exception of some isolated points (energies). Alternatively we can have to do with decoherence by inelastic scattering. Metallic regime was also measured on 2D electron systems in Si MOSFETs \cite{akea}, two parameter scaling involved, and it was again explained by above mentioned e-e interactions. Thus the single parameter scaling theory predictions, especially no MIT in orthogonal 2D systems, are disputed from various sides. The non-satisfactory situation was summarized in \cite{aks}, stating that there is no simple model capable to explain all experimental data at once.

Superlocalization, i. e. over-exponential decay of $g(L) \propto \exp{-(L/\xi)^\kappa}, \kappa > 1$, was shown in Ref. \cite{khs} on the percolation cluster at $p=p_c$ and explained as a consequence of the cluster's fractal nature. Here $p$ is the ratio of randomly removed sites and $p_c$ is the percolation threshold. We do not reach the classical 2D site percolation threshold $p_c=0.407$. 

The charge careers, represented by planar quantum waves, have the energy $E$. The conduction band for simple hypercubic sample with the volume $L^d$ is the interval $-2d < E < 2d$. Anomalous behavior, meaning non-exponential decay of $g(L)$, was reported \cite{kvn} mainly at band center, i. e. for small energy $|E|$. In Ref. \cite{kvn} Thouless conductivity $\sigma$ was calculated from the dependence of energy spectra on boundary conditions and a transition between two types of localization was reported even in 2D. One expected type should be sublocalization or weak localization, i. e. power-law decay of $\sigma(L)$ with negative exponent. We will test such possible behavior for Landauer conductance $g$ and the largest $L$ available.

In localized regime, the energy dependences of conductance have resonances, i. e. sharp peaks. They ofted reach to conducting values, i. e. $g_{max} \approx 2$. The nature of wave functions at resonant energies \cite{dsm} has shown that they are of Pendry necklace type, i. e. a chain of comparable localized wave packets. Such a quasi 1D chain enables the transmission across the sample also in higher dimensional case. Pendry predicted \cite{pen}, that even for $d>1$ the probability $r$ of these rare states should decay with $L$ as $r \propto \exp(-\sqrt{L/\xi})$, a relation he derived analytically in 1D.

As the main tasks, in this paper we aim to look for signs of MIT or sublocalization in the whole conduction band, to test Pendry's conjecture on rare events and to present the properties of the quasi-metallic regime, compared even with quantitative predictions.

\section{Model and Method}

Let us now introduce the tight-binding Hamiltonian:

\begin{equation}
\label{eq:1}
H = \sum_i \epsilon_i |i><i| + V \sum_{<i,j>} |i><j|
\end{equation}

\noindent where $i$ numbers the $L\times L$ lattice sites and the second sum goes over nearest neighbors. Later we set the hopping $V=1$ in arbitrary units, thus fixing the energy scale. With box distribution of random on-site energies $P(\epsilon_i)$, $ - W/2 < \epsilon_i < W/2$, the Hamiltonian (\ref{eq:1}) becomes the Anderson model and $W$ is called disorder. For a binary system the distribution of the energies has the form:

\begin{equation}
\label{eq:2}
P(\epsilon_i) = (1-p) \delta(\epsilon_i-\epsilon_A) + p \delta (\epsilon_i-\epsilon_B),
\end{equation}

\noindent where $p$ is the ratio of insulating sites $B$ and it is the measure of disorder in this model. $\delta(x)$
is Dirac's delta function. We can analyze any combination of energies $\epsilon_A$ and $\epsilon_B$ by our method, but in this paper we concentrate on the quantum percolation model. Hence we set the on-site energy for A-type atoms to $\epsilon_A = 0$ and the one for the randomly positioned B-type atoms to a large number, mostly $\epsilon_B= 10^7$. For a completely removed, inaccessible site this energy would be infinite. The difference will be discussed in the next chapter. We chose this approach in order to apply the transfer matrix method to calculate the transport properties. When we study the eigenenergies, namely their density of states (DOS) $\rho$, we simply set all on-site energies to zero and also the hopping elements $V_{ij}=0$ for all nearest neighbors $j$ of the excluded sites $i$.

We use standard transfer matrix method for numerical calculations. The square sample is connected to two perfect semi-infinite leads. A set of $L$ (generally $L^{d-1}$) planar waves enters the square sample from the left. They have different wavenumbers in transfer direction $k_n$, $n=1, ..., L$, given implicitly by the dispersion relation

\begin{equation}
\label{eq:3}
E = -2 V \cos k_n a -2 V \cos {\frac{\pi n a}{L+1}}, \ \ \ \ 0 \le k_n \le \pi/a.
\end{equation}

We set the lattice constant $a=1$, thus fixing the length scale. The last term is given by hard wall boundary conditions in perpendicular direction. The real solutions for $k_n$ are called open channels and they are possible only for $|E|< 4$; the complex solutions lead to exponentially damped waves - closed channels and they are omitted at the end. The part of the waves arriving on the right hand side of the sample is described by the transmission matrix $t$, the rest was reflected to the left and it is described by the reflection matrix $r$, both $L\times L$. The elements of the transmission matrix $t_{\alpha\beta}$ are the amplitudes of charge careers transmitting from channel $\alpha$ (on the left) to the channel $\beta$ on the right. The conservation law has the form $t^\dagger t + r^\dagger r = ${\bf 1}, where {\bf 1} is identity matrix representing incoming waves and this is a generalized Kirchhoff law. The conductance $g$ in units $e^2 / h$ is calculated making use of Landauer formula $g= 2\ {\rm Tr}\ t^\dagger t$, the trace restricted to open channels. The $2L\times 2L$ transfer matrix $T$ relates waves on the right: vector $V_r$ of $L$ outgoing transmitted waves and $L$ (zeros) incoming ones, to those on the left: $V_l$ of $L$ incoming and $L$ reflected waves: $V_r = T V_l$. It can be shown to fulfill \cite{pmr}

\begin{equation}
\label{eq:4}
T = \begin{pmatrix}
t^{-1} & -t^{-1} r \\
r^\dagger t^{-1} & t^\dagger - r^\dagger t^{-1} r \\
\end{pmatrix}
\end{equation}

The main advantage of this formulation is the possibility to cut the sample into slabs, perpendicular to the transfer direction. The $L$ transfer matrices of these slabs are then multiplied to get the complete $T$, because outgoing waves of one slab are the incoming ones of the next slab. 
But we have to do with ill-defined matrices, i. e. their eigenvalues differ by many orders of magnitude for samples even much smaller than it was the case for Anderson model. In other words, numerical errors, given mainly by closed channels, rise quickly with sample size. An efficient way to treat this problem was described in Ref. \cite{pmr} and at the end of Ref. \cite{pma}. It is based on the fact, that the upper matrices involved in Eq. (\ref{eq:4}) have large eigenvalues dominating $t^{-1}$, but we need accurate lower eigenvalues to recover $t$. Thus we multiply the $L \times L$ submatrices analogous to those in Eq. (\ref{eq:4}) by the inverse of their top half and store the product of changes made by this to be multiplied with $t$ at the end. In quantum percolation, we have to perform the procedure at least after each two transfer matrix multiplications (cca 10 in Anderson model), but then we are able to include larger samples up to $L=180$. The most numerically sensitive region is the anomalous one, where maximum $L \approx 100$ for $p\ge 0.1$.

A slight drawback of our method is, that even the large chosen $\epsilon_B$ still allows small conductance $g$ via quantum tunneling - we can guess this value roughly by constructing a single perpendicular barrier of thickness 1 site, and calculate its $g \approx 10^{-11}$ for sample sizes around $L=100$ (and $\ln g \approx -25$). For two distant barriers we get $g \approx 10^{-21}$ after tunneling twice. Deep insulators must be treated
carefully, in order to distinguish between Anderson localization and quantum tunneling.

A part of our finite samples creates percolating cluster of insulating sites even for ratios of these sites $p < p_c$, recall that $p_c$ is the classical site percolation threshold. In classical percolation such a sample would have zero conductivity, in our case quantum tunneling gives small non-zero conductance of above mentioned orders. Neither taking these samples into account nor omitting them is fully correct. Thus when we calculate $g$ in the localized regime, we prefer large samples and ratios $p$ not quite close to $p_c$. The accuracy of $\langle \ln g\rangle$ is somewhat spoiled and we will therefore use the less sensitive $\langle g\rangle$ and $var\ g = \langle g^2\rangle - \langle g\rangle^2$, if possible. At least this is easy to test by changing $\epsilon_B$. If the conductance changes, we have tunneling or another (numerical) problem. Otherwise we have Anderson localization and the data are plausible.

\bigskip
\bigskip

\begin{figure}[htb]
\begin{minipage}[t]{.43\textwidth}
\includegraphics[width=\textwidth]{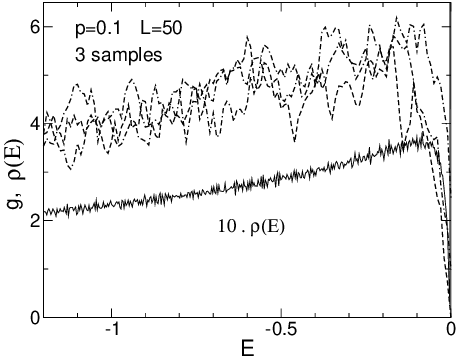}
\caption{The conductance $g$ vs. energy in metallic regime and the DOS $\rho(E)$, full line.}
\label{fig:gef}
\end{minipage}
\hfil
\begin{minipage}[t]{.43\textwidth}
\includegraphics[width=\textwidth]{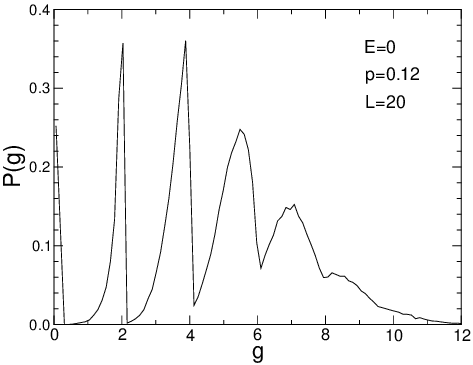}
\caption{The partially quantized distribution of $g$ in the band center $E=0$.}
\label{fig:pgq}
\end{minipage}
\end{figure}

\section{Numerical results}

At first let us describe the so called anomaly region at the center of the conduction band, i. e. for small energy $|E|$. It is known, that $E=0$ can be a special case in these models, e. g. in 1D Anderson model the wave function gains additional symmetries, Ref. \cite {dela}. Fig. \ref{fig:gef} shows the conductance as a function of energy, $g(E)$, for three realizations of positions for B-type sites, with the same ratio $p$. We can see that, outside of the center, $\langle g\rangle$ slightly increases with the number of open channels (which itself decreases with $|E|$) and $g$ oscillates with approximately constant fluctuations. The ergodic theorem states that these fluctuations of $g$ are the same as those for many different samples with the same $p$ and $E$, called statistical ensemble. We will calculate these fluctuations in the latter way. But near the band center $g$ quickly decreases to some random value, much smaller than $g$ far from center. Such behavior remains in localized regime, too. This anomaly is not present in the 2D Anderson model. We believe it is connected with the decreasing density of states (DOS) $\rho$ at the band center\cite{kvn, ba}, because the ensemble averaged DOS ($10 \times$ scaled for convenience) $\rho(E)$ is similar to $\langle g(E)\rangle$ in metallic regime, see Fig \ref{fig:gef}. The lack of eigenstates reduces the transmission, though not in a trivial manner. We found that mean $g(E;p,L)$ is not simply proportional to $\rho(E;p,L)$. An analogical graph, without DOS, was published in Ref. \cite{zs}, where even an average over 100 configurations did not remove the fluctuations completely, not to speak about the anomalous decay. Similar DOS, with larger disorders $p$, where presented in Ref. \cite{kvn}. The high peak at $E=0$ was removed.

Another interesting feature of the anomaly is the partial quantization of conductances, see Fig. \ref{fig:pgq}. It was already reported in Ref. \cite{al} for $g$ as a function of the wave number $k$. Here it happens at the band center, for medium disorder $p$ and it is especially well pronounced for smaller samples. The most typical values of $g$ are slightly below 2, or 4, etc. This means that we have 1, or 2 conducting channel(s) and the rest is almost negligible, as the peaks are highly asymmetric, practically no 2+, or 4+ conductances. For larger $g$ the peaks are shifted to lower values (lower than 6, 8) and suppressed, again as in Ref. \cite{al}, where it was shown that the quantization disappears in the thermodynamic limit. Accordingly in our case both for larger $|E|$ and/or $L$ this quantization is gradually smeared out. Closer inspection of samples shows that these conducting channels are not trivially connected to existing straight lines of conducting sites across the sample, because ``pathways'' with short detours can also give $g \approx 2$ and others can have very small $g$.

\begin{figure}[htb]
\includegraphics[width=.39\textwidth]{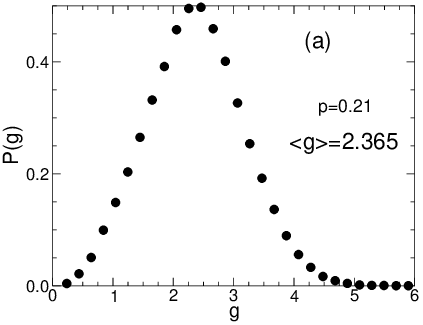}~(a)
\hfil
\includegraphics[width=.39\textwidth]{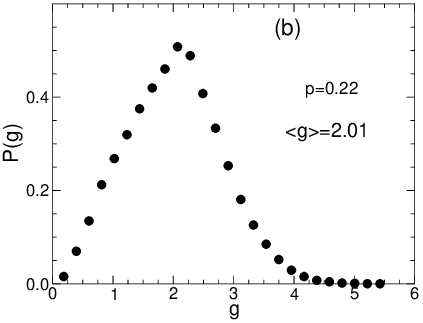}~(b)
\vskip0.4truecm
\bigskip
\includegraphics[width=.39\textwidth]{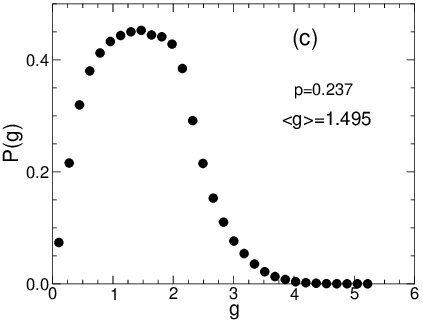}~(c)
\caption{$P(g)$ for $E=-0.53$, $L=80$ and various $p$.}
\label{fig:pg}
\end{figure}

There is no smooth change of $P(g)$ with one maximum from Gaussian distribution in metal to deformed log-normal distribution in the insulator, as it is the case far from band center, Fig. \ref{fig:pg}, and for the Anderson model. Instead a part of samples enters these quantized peaks and the others enter the clearly separated peak of localized samples close to $g=0$. With larger $L$ more samples become localized and less samples remain quantized, regardless of (fixed) $p$. We found no critical $P_c(g)$, that would be $L$ independent. This is why we are convinced that there is no MIT even at $E=0$ and the discontinuous behavior of conductances could have lead some authors to different statements, e. g. Ref. \cite{dca}.

\bigskip
\bigskip

\begin{figure}[htb]
\begin{minipage}[t]{.43\textwidth}
\includegraphics[width=\textwidth]{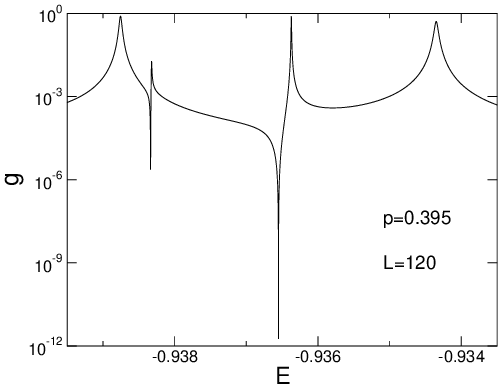}
\caption{Resonance peaks of $g(E)$ for insulator.}
\label{fig:res}
\end{minipage}
\hfil
\end{figure}

Further let us address the problem of resonances, manifest in the deeply localized regime in the whole conduction band. For smaller samples we get the well known resonance peaks with relatively high amplitudes yielding maxima close to the value $g_{max} = 2$, see e. g. Ref. \cite{mds}.

We present $g(E)$ semilogarithmically on a small interval of $E$ with several typical peaks in Fig. \ref{fig:res}, the maxima being roughly Lorentzians. We often get a coupled minimum - maximum pair, apparent in $\ln g(E)$ plot. These line shapes are somewhat reminiscent of those from Azbel - Kaner \cite{ak} cyclotron resonance, e. g. Ref. \cite{ch}; just the latter are for $g$ instead of $\ln g$. Of course, we have no successive maxima $E_n=n E_1$, as we have no magnetic field and the similarity is probably occasional. There are also few isolated minima, even deeper than the paired ones. The minima where not reported yet, as they are not very apparent in linear scale. 

The values of global maxima are slightly below 2, say $g \approx 1.95 - 1.99$, in analogy to the quantization in the anomalous region. This means that we have one open channel, almost perfectly conducting at special tuned energy and the contribution of other channels can be neglected at maxima. Calculating an ensemble of $10^5$ samples with fixed $E$ and $p$ up to 0.4, virtually never gives conducting samples. Nevertheless performing a rough energy scan of one sample and then tuning the $E$ of chosen maximum up to 9 digits allows to find them already within few samples - this is a much more effective strategy for finding these rare events. Contrary to anomalous quantization we never see $g_{max} \approx 4$, i. e. two channels tuned to same $E$. The peaks are homogeneously spread along the whole conduction band.

Now let us switch to more common strategy of statistical emnsembles. I. e. we calculate the properties of many samples with fixed $E$ and $p$, just not too close to $p_c$. For $E=-0.53$, i. e. far from band center, medium system sizes $L$ and large ensembles ($10^6$ samples) we fitted the ratio of extended cases, Ref. \cite{pen}, $r(g>2) \propto \exp(-c L^\kappa)$ and got the value $\kappa = 0.52(11)$, close to Pendry's $1/2$. We performed identical analysis for Anderson model and got almost the same $\kappa \approx 0.55(12)$. Error bars are rather large, the $\ln r(L)$ data get noisy for very small $r$ and this reason disabled analogous verifications in 3D cases. Nevertheless these results are strong support of Pendry's conjecture $\kappa = 1/2$ for both models in 2D. The $g=2$ is a natural turning point as, apart from the anomalous band center, $g>2$ usually means, that at least one or two channels have transmission close to 1 and $g<2$ arises typically as a sum of many channels with low transmission; this is the well-known cause of non-analyticity of $P(g)$, Ref. \cite{rms}, see Fig. \ref{fig:pg} b, c. Contrary to the anomalous region, the quantized cases with $g \approx 1.99$, become very rare within the ensemble approach for deep localization case, as the peaks in Fig. \ref{fig:res} are narrow.

\bigskip
\bigskip
\bigskip

\subsection{Distributions of conductance, far from band center}

The distribution $P(g)$ is Gaussian in metallic regime, but in crossover regime between metal and insulator, $L \approx \xi$, it differs significantly from Anderson model's distributions even outside the anomaly region. It strongly resembles the distributions calculated in Ref. \cite{gms} for quasi-1D wires with rough surfaces. It is interesting, that assembling (still randomly) the insulating sites on one side or dispersing them in volume does not change typical $P(g)$ even in the crossover regimes. We can confirm that also in our model $P(g)$ can be categorized by $\langle g\rangle$ and it depends only slightly on the combination of other parameters, giving the same $\langle g\rangle$, thus confirming the single parameter scaling. Just our $P(g)$ of the same shape correspond to somewhat lower $\langle g\rangle$ values than theirs (Ref. \cite{gms}, beware the units), but our samples are much larger. We recall that $E=-0.53$ was now chosen so that we are far from the anomalous band center and $P(g)$ changes continuously with $p$, i. e. its maximum shifts gradually instead of being fixed to some quantized values as in Fig.\ref{fig:pgq}.

Some noisy (because of much smaller statistics) $P(g)$ histograms for our model were already published \cite{zs}. We used ensembles of 200.000 samples.

As usual in these models, we have non-analyticity of $P(g)$ at $g=2$. The only difference in shapes is, compared to wires with rough surfaces, that our $P(g)$ in Fig.\ref{fig:pg}a is less sharp than that in Ref. \cite{gms}; it turns gradually to a Gaussian and its maximum is slightly above $g=2$; for $p$ even smaller the peak shifts towards larger $\langle g \rangle$. The sharper peaks in \cite{gms} were calculated for very small number of open channels, which is not our case.

\subsection{Cumulants of conductance}

\begin{figure}[htb]
\begin{minipage}[t]{.43\textwidth}
\includegraphics[width=\textwidth]{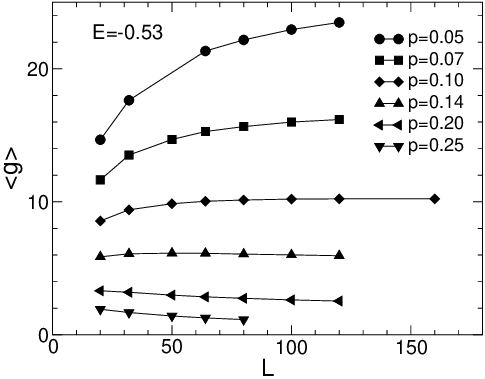}
\caption{Mean conductance as a function of system size far from band center.}
\label{fig:gl}
\end{minipage}
\hfil
\begin{minipage}[t]{.43\textwidth}
\includegraphics[width=\textwidth]{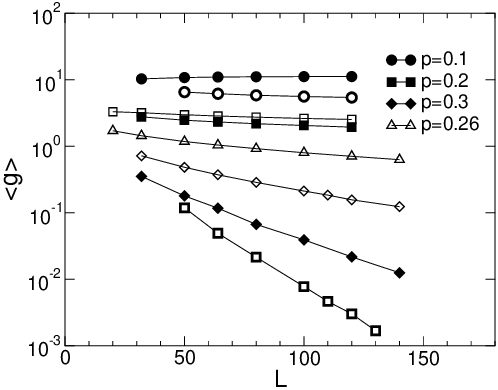}
\caption{Mean conductance vs. $L$. Symbols shapes define $p$. Filling defines $E$, see text.}
\label{fig:gll}
\end{minipage}
\end{figure}

Let us now plot several dependences of $\langle g\rangle$ and $var\ g$ on model parameters $L$ or $p$. In analogy to Anderson model we expect that the leading term of the mean conductance $g \propto L$ in the ballistic regime (for both $L$ and $p$ very small) and $g \propto L^0$ in the metallic - diffusive regime. In Fig.\ref{fig:gl} we can see that indeed for $p \approx 0.1$ we get a wide region where $\langle g\rangle$ is almost independent of available $L$, though we expect that for any positive $p$, large enough $L$ (beyond computing possibilities) would localize all modes and there is no MIT.
To confirm this, we calculated up to $L=160$ and found a small but significant decrease, the onset of localization.

\bigskip
\bigskip
\bigskip

\begin{figure}[htb]
\begin{minipage}[t]{.43\textwidth}
\includegraphics[width=\textwidth]{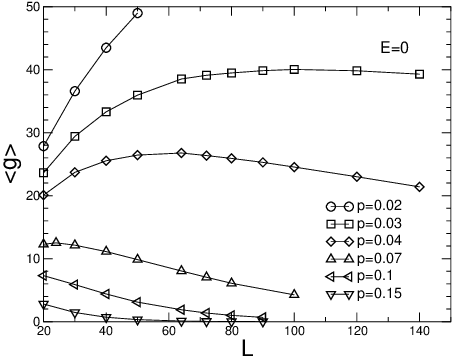}
\caption{Mean conductance as a function of system size in the band center.}
\label{fig:gle0}
\end{minipage}
\hfil
\begin{minipage}[t]{.43\textwidth}
\includegraphics[width=\textwidth]{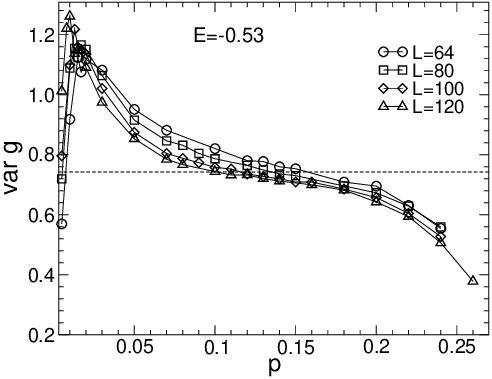}
\caption{Conductance fluctuations vs. disorder for various $L$. Broken line is theoretical UCF.}
\label{fig:ucf}
\end{minipage}
\end{figure}

In the insulating regime, there were indications, that the localization in the anomaly region could be non-exponential, e. g. power law or superlocalizing.
It was already shown in Ref. \cite{zs} numerically, that at least in 2D for $L$ large this is not the case - the $\langle \ln g\rangle$ vs. $L$ plot is linear also for small $|E|$. We show in bottom part of Fig.\ref{fig:gll} that even $\langle g(L)\rangle$ in semilogarithmic scale is almost linear for larger $L$, thus the Anderson localization is present both inside ($E=-0.02$, symbols with broad boundary) and outside the anomaly region ($E=-0.2$, full symbols, and $E=-0.53$, empty ones). The deviation from linearity at smaller samples, see e. g. the lowest line, could be misleading. We present here also $\langle g(L)\rangle$ in the anomalous band center $E=0$, Fig.\ref{fig:gle0}. It is similar to Fig.\ref{fig:gll}, but the plateau-like regions, i. e. (quasi)metallic parts around maxima are shorter and shifted to lower disorders $p$ and higher $\langle g \rangle$. We can see exponential decay, i. e. Anderson localization, for $p>1$ and thus no MIT for $p=0.26$ and $E=0$, reported in Ref. \cite{dca}.

We will further analyze the conductance fluctuations. In metallic regime the half width of $P(g)$ Gaussians is an universal constant, depending only on physical symmetries, boundary conditions (BC) and dimensionality, but independent of energy, system size and disorder - for medium $p$ and $L$; in higher dimensionalities $d>2$ even for the largest $L$ available. We use the hard wall BC in perpendicular direction, quantizing the last term in Eq. (\ref{eq:3}). The slow dependence of $\sqrt{var\ g}$ on $L$ was already published \cite{zs}, we prefer $var\ g(p)$, where the theoretical prediction approaches the inflex point, see Fig.\ref{fig:ucf}.

\bigskip
\bigskip

The universal conductance fluctuations (UCF) predicted for 2D hard wall case ought to be $var\ g \approx 4 * 0.1856 = 0.7424$ in our units, e. g. Ref. \cite{lsf, it}. One can see in Fig.\ref{fig:ucf} that after the ballistic
peak there is a region, where $var\ g$ only very slightly decreases with $p$
and this happens around the theoretically predicted value. This graph is very similar to the one calculated for Anderson model \cite{rms2}. This is true only outside the anomaly region; $var\ g$ at maxima in Fig. \ref{fig:gle0} become non-universal and at least by one order of magnitude greater than UCF - in band center we are still in ballistic regime. Nevertheless, the metallic regime is not just the only one exactly solved yet, predicting Gaussian $P(g)$ with known universal halfwidth and $\langle g\rangle \propto L^{d-2}$; it is also the most robust regime for the mutually related models with disorder-induced localization.


\section{Conclusions}
We have shown, that the transfer matrix method suits well for studies of the 2D quantum percolation model as a limit of binary alloy. The regions of interest can be divided according to following two criteria. The first one is the regime of transport, given mainly by disorder $p$: ballistic, quasi-metallic, crossover, localized and deeply localized regimes. Another one is according to energies $E$: close to band center (small $|E|$, called anomalous region) and far from it. 

The statements of MIT present in 2D quantum percolation model, Refs. \cite{dca} and \cite{im}, were achieved with 1D linear leads and smaller conduction band than ours. We have leads of thickness $L>1$. It was shown in Ref. \cite{sop} that narrow leads give different results, at least in 2D Anderson model, than the wide ones. On contrary in Ref. \cite{hks} the absence of MIT was shown for $E=0.5$. We did not find the MIT either, even at $E=0$. Increasing $g(L)$, that would give rise to positive values of the scaling function $\beta (g) = d \ln g / d \ln L$ (and MIT at $\beta (g_c)= 0$), appears to be just finite-size effect. We believe that $\beta(g)$ is always negative\cite{hks}. The anomaly region near the center of the conduction band retains Anderson localization in large samples and we found neither sublocalization (power-law decay of $g(L)$), nor superlocalization (over-exponential) for large $L$ at any $E, p$. Instead, deviations appear that are again just finite-size effects, unusually magnified by the novel partial quantization of $g$. It is a question, whether we should even call the conducting regime at band center (quasi)metallic, as the plateau of mean $g(L)$ is too small, no UCF are observed and the onset of exponential decay (localization) appears already for $g>2$. The only reason left is that the peaks of $P(g)$ apart from quantization region, say those with $\langle g \rangle > 10$, are almost Gaussian. Anomalous behavior, seen theoretically in 1D at $E=0$ and sometimes conjectured to higher dimensionalities, is not found numerically for Anderson model in 2D and 3D, but it appears in the quantum percolation model for small and medium $L$. We suppose that the decay of $g(E)$ at band center is also connected with the decrease of DOS in the sense that effectively non-zero $\rho$ is a necessary condition for non-zero $g$. 

Most Lorentzian (or Cauchy) resonance peaks in the strongly localized case are of comparable relative height in semilog-scale and thus they do not make the samples always conductive in our calculations of deep insulators, i. e. for both $p$ and $L$ large. Pairs of minimum and maximum - or vice versa - occur in this scale that were not reported yet. 

The probability $r(g>2)\propto \exp(-c L^\kappa)$ of rare delocalized samples confirms Pendry's conjecture \cite{pen} $\kappa = 1/2$ both for 2D quantum percolation far from band center and for 2D Anderson model. This is the first numerical confirmation of the conjecture. The rare, i. e. conducting cases within deep insulators, close to classical percolation threshold, can be found by tuning the energy at maxima in $g(E)$ plots even for medium sized samples. 

The distributions of conductance in the crossover regime far from band center are very similar to those of quasi-1D wires with rough surfaces and they are different from those of Anderson model. In the metallic phase the mean conductance is practically $L$ independent in a large region and, except band center, its variance is close to the universal (UCF) value predicted theoretically \cite{lsf}.

Concludingly let us compare the Anderson model and the quantum percolation one. We are convinced that they belong to the same universality class, i. e. they undergo no MIT for $d \le 2$. Running forward, in our most recent paper \cite{it2} we show that the models have identical critical exponent in 3D MIT. Back in 2D, far from band center they behave very similarly, apart of tiny differences in crossover regime. At band center, quantum percolation model has very strong finite-size effects, somewhat questioning even the single parameter scaling hypothesis. Such effects are not present in Anderson model.

This work was supported by Grant VEGA Nr. 2/6069/26.


\end{document}